# Autonomous Orbit Determination via Kalman Filtering of Gravity Gradients

Xiucong Sun, Pei Chen, Christophe Macabiau, and Chao Han

*Abstract*—Spaceborne gravity gradients are proposed in this paper to provide autonomous orbit determination capabilities for near Earth satellites. The gravity gradients contain useful position information which can be extracted by matching the observations with a precise gravity model. The extended Kalman filter is investigated as the principal estimator. The stochastic model of orbital motion, the measurement equation and the model configuration are discussed for the filter design. An augmented state filter is also developed to deal with unknown significant measurement biases. Simulations are conducted to analyze the effects of initial errors, data-sampling periods, orbital heights, attitude and gradiometer noise levels, and measurement biases. Results show that the filter performs well with additive white noise observation errors. Degraded observability for the along-track position is found for the augmented state filter. Real flight data from the GOCE satellite are used to test the algorithm. Radial and cross-track position errors of less than 100 m have been achieved.

## I. Introduction

GEOPHYSICAL information, such as the Earth's gravity and magnetic fields, is of particular interest for fully autonomous or GPS-denied navigation [1]. Spaceborne magnetometers which provide magnetic field measurements, for example, have been proposed for autonomous spacecraft orbit and attitude determination. By matching the observations with the International Geomagnetic Reference Field (IGRF) model, position errors of several kilometers and attitude errors from 0.1 to 5 deg have been achieved with simulated and real fight data [2-5]. The gravity gradiometer, which can sense full-tensor or partial-tensor gravity gradients, has also been pursued as a means of navigation. The GPS signals might be jammed or spoofed by ground-based or space-based attackers [6]. By contrast, the gravity gradient signals could not be interfered easily. In addition, the GPS navigation is available to Earth orbits only and is ineffective for exploration missions far from the Earth, such as the Moon or Mars. The gravity gradient based navigation does not rely on ground stations or any other satellites and is an ideal choice for autonomous spacecraft operation as well as for envisioned future exploration of Earth-like planets.

This research was supported by the National Nature Science Foundation of China through cooperative agreement No. 11002008 and has been funded in part by Founded by MOST through cooperative agreement No. 2014CB845303. The European Space Agency is acknowledged for providing the GOCE data.

X. Sun, P. Chen, and C. Han, School of Astronautics, Beihang University, 37 Xueyuan Road, Beijing, 100191, People's Republic of China (e-mail: chenpei@buaa.edu.cn).
C. Macabiau, TELECOM Lab, Ecole Nationale de l'Aviation Civile, 7 Avenue Edouard Belin, 31055 Toulouse Cedex 04, France.

In early researches the gradiometer was considered as a useful aid to the inertial navigation system (INS) [7-9]. Two integration methods, gravity disturbance compensation and gravity gradient map matching, were studied. The former made use of gravity gradients combined with the velocity information to estimate gravity disturbance forces, whereas the latter utilized a stored map for observation data matching and then to estimate the trajectory directly. In 1990, Affleck and Jircitano [10] presented a passive navigation system, in which the measured gravity gradients were compared with map values and the differences were processed by an optimal filter to correct INS errors. A parametric study was conducted for an airborne INS. Position accuracies of better than 100 m were demonstrated. Gleason [11] continued the work and developed a fast Fourier transform algorithm to efficiently generate constant altitude grids of reference gravity gradients. More recently, Richeson [12] presented a comprehensive discussion on the gravity gradient map matching technique for inertial navigation aiding. Monte Carlo simulations of a hypersonic cruise showed that sub-meter position errors could be possible with a future grade gradiometer.

Over the past few decades, gravity models of terrestrial planets, especially the Earth, have been improved dramatically. Derived from the combination of satellite geodetic data with high-resolution gravitational information collected from surface gravimetry, the new developed Earth Gravitational Model 2008 (EGM2008) is complete to degree 2190 and order 2159 [13]. Meanwhile, significant advances in spaceborne gravity gradiometry have occurred due to geophysical activities [14]. The electrostatic gravity gradiometer (EGG) implemented on ESA's gravity field and steady-state ocean circulation explorer (GOCE) achieved a noise density level of 0.01 E/√Hz within measurement bandwidth (MBW). A further better accuracy of $10^{-3}$ E/√Hz is projected in the future using cold atom interferometers [15].

The progresses in global gravity field modeling and spaceborne gravity gradiometry provide opportunities for the application of gravity gradiometry to spacecraft navigation. Chen [16] introduced an idea of using full-tensor gravity gradients combined with precise inertial attitudes to determine position. The system's observability was explained from geometry and the effects of possible sources of errors were analyzed. An eigendecompostion based positioning algorithm using the $J_2$ gravity model was developed. Simulation results showed that a mean position error of 421 m could be possible for a spacecraft with a gradiometer having a noise level of 0.1 E at 300 km altitude. Sun [17] investigated the use of a least squares searching method which employed a high-degree gravity model for map matching. Real flight data from the GOCE satellite were used to test and verify the method and a mean positioning error of 620 m was achieved.

In this study, autonomous orbit determination using gravity gradients within an extended Kalman filter (EKF) is presented. The main advantage of the EKF-based orbit determination is that the measurement noise can be reduced through incorporation of the orbital motion. Moreover, spacecraft velocity can be output simultaneously. Within the EKF, a 120th degree and order EGM2008 model is used to calculate gravity gradients and high-precision inertial attitudes are used to remove the contribution of instrumental orientation. An augmented state EKF is also developed to deal with significant measurement biases. The algorithms are applied to both simulated data and real GOCE data and effects of several important factors are analyzed.

The remainder of this paper is organized as follows. Section II introduces the mathematical model of gravity gradients. Section III presents the EKF-based orbit determination algorithm, including orbital dynamic modeling, the measurement equation, the estimation formulas, and the model configuration. Section IV presents the metrics used for evaluating filter performance. Simulation results are given in Section V. The effects of initial errors, data-sampling periods, orbital heights, noise levels, and measurement biases are discussed. Section VI describes the GOCE satellite whose flight data have been used to test the algorithm, and the orbit determination results are presented. The major finding of this study is concluded in Section VII.

## II. Gravity Gradient Modeling

The gravity gradients (or gravitational gradients) are the second-order spatial derivatives of the gravitational potential and form a 3 × 3 matrix called the gravity gradient tensor (GGT)

$$V = \nabla(\nabla U) = \begin{bmatrix} \frac{\partial^2 U}{\partial x^2} & \frac{\partial^2 U}{\partial x \partial y} & \frac{\partial^2 U}{\partial x \partial z} \\ \frac{\partial^2 U}{\partial y \partial x} & \frac{\partial^2 U}{\partial y^2} & \frac{\partial^2 U}{\partial y \partial z} \\ \frac{\partial^2 U}{\partial z \partial x} & \frac{\partial^2 U}{\partial z \partial y} & \frac{\partial^2 U}{\partial z^2} \end{bmatrix} = \begin{bmatrix} V_{xx} & V_{xy} & V_{xz} \\ V_{yx} & V_{yy} & V_{yz} \\ V_{zx} & V_{zy} & V_{zz} \end{bmatrix} \quad (1)$$

The unit of GGT is Eötvös, denoted by the symbol E. One Eötvös equals $10^{-9}$ $1/s^2$ in SI units. There are only five independent terms in matrix $V$. The continuity of the gravitational field guarantees that $V$ is symmetric and the Laplace's equation constrains its trace to be zero [18].

The expression of GGT depends on the choice of reference system. Expressions of GGT in two different reference systems have the following relationship

$$\begin{aligned} V_b &= C_a^b V_a \left(C_a^b\right)^T \\ &= C_a^b V_a C_b^a \end{aligned} \quad (2)$$

where the symbols $a$ and $b$ denote the reference frames, the superscript $T$ denotes the transpose operation, and $C_a^b$ is the transformation matrix from frame $a$ to $b$.

The common reference frame used in most of the Earth gravity models (EGM96, EGM2008, GGM03, etc.) is the Earth-centered Earth-fixed (ECEF) frame, one realization of which is the International Terrestrial Reference Frame (ITRF) developed by the International Earth Rotation and Reference Systems Service (IERS). The ECEF frame cannot be physically materialized onboard a satellite. The gravity gradients can only be measured in the Gradiometer Reference Frame (GRF), which is defined by the three orthogonal baselines of a gradiometer. The transformation matrix from ECEF to GRF will be specified in Section III for the measurement equation formulation.

A series of spherical harmonics is used in geodesy to model the gravitational potential

$$U(r,\phi,\lambda) = \frac{GM}{r}\sum_{n=0}^{\infty}\left(\frac{R_E}{r}\right)^n \sum_{m=0}^{n} \overline{P}_{nm}(\sin\phi)\left[\overline{C}_{nm}\cos(m\lambda) + \overline{S}_{nm}\sin(m\lambda)\right] \quad (3)$$

where $r$, $\phi$, and $\lambda$ are the geocentric distance, latitude, and longitude of the position (expressed in ECEF), $GM$ is the geocentric gravitational constant, $R_E$ is the reference equatorial radius of the Earth, $n$ and $m$ are the degree and order of the normalized coefficients $\overline{C}_{nm}$ and $\overline{S}_{nm}$, and $\overline{P}_{nm}$ is the normalized Legendre function. The values of $GM$ as well as the coefficients, $\overline{C}_{nm}$ and $\overline{S}_{nm}$, are determined from observations such as satellite tracking and surface gravimetry. The gravity gradients can be obtained by double differentiation of Eq. (3). The calculation formulas are given in Appendix.

The spherical harmonic model is usually truncated at a maximum degree for the practical computation. The precision of gravity models with respect to GGT truncated at different degrees is shown in Fig. 1, in which the 300th degree and order EGM2008 model is used as a benchmark. The reference height is 300 km. It can be seen that the mean GGT error decreases with increase of the truncating degree. The effects of tidal variations (solid Earth tide, ocean tide, and pole tide) as well as the third body attraction (Sun and Moon) on GGT modeling are insignificant and are not considered.

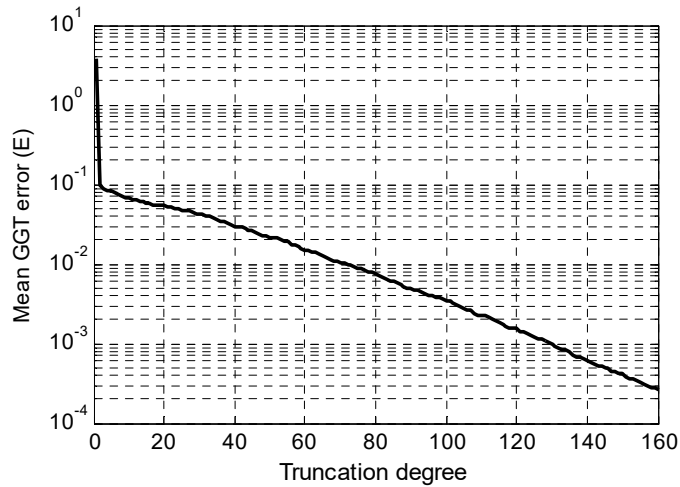

Fig. 1. The precision of EGM2008 gravity models truncated at different degrees. The benchmark is a 300th degree and order version. The reference height is 300 km.

III. ORBIT DETERMINATION ALGORITHM

Orbital motion of spacecraft is subject to various random perturbations. The orbit determination problem is to estimate the states (position and velocity) of such a stochastic dynamic system from noisy observations, which in this study refers to the gravity gradient measurements. The EKF is a suboptimal filter which gives real-time estimates of the orbital states and the error covariances based on sequential observations. This section describes the dynamic model and the measurement equation and presents details of the estimation algorithm.

*A. Stochastic Model of Orbital Motion*

The orbital motion is modeled as a nonlinear stochastic dynamic system and is described by the following stochastic differential equation

$$\frac{d}{dt}\begin{bmatrix} r \\ v \end{bmatrix} = \begin{bmatrix} v \\ f(r,v,t) \end{bmatrix} + \begin{bmatrix} 0 \\ w_t \end{bmatrix} \quad (4)$$

where the state vector comprises the position and velocity ($r$ and $v$) in the ECI frame, $f$ is a 3-dimensional vector function representing the accelerations due to the deterministic forces, and $w_t$ represents the remaining unmodeled perturbation accelerations and is assumed to be a zero-mean stationary white Gaussian noise.

The continuous time dynamic model needs to be discretized before being used for the state prediction in the EKF, since the measurements are taken at discrete instants. The discretized dynamic equation has the following form

$$x_k = \varphi(x_{k-1}, t_k, t_{k-1}) + w_{k-1} \quad (5)$$

where $x_k$ is the orbital state at time $t_k$, $\varphi$ is a 6-dimensional vector function relating states at two adjacent instants, and $w_{k-1}$ is the discrete process noise. The random sequence $\{w_k\}$ is independent and zero-mean Gaussian. The function $\varphi$ has no explicit expression and is only numerically obtained using an ordinary differential equation (ODE) solver.

The first-order partial derivative of $x_k$ with respect to $x_{k-1}$ is called the state transition matrix, denoted as $\Phi(t_k, t_{k-1})$. $\Phi(t_k, t_{k-1})$ can be obtained by numerically integrating the following differential equation from $t_{k-1}$ to $t_k$

$$\frac{d}{dt}\Phi(t_{k-1}+t, t_{k-1}) = \begin{bmatrix} 0_{3\times3} & I_{3\times3} \\ \left(\frac{\partial f}{\partial r}\right)_{3\times3} & 0_{3\times3} \end{bmatrix} \Phi(t_{k-1}+t, t_{k-1}) \quad (6)$$

with the identity matrix as the initial value

$$\Phi(t_{k-1}, t_{k-1}) = I_{3\times3} \quad (7)$$

The covariance matrix of $w_{k-1}$ is called the process noise matrix, denoted as $Q_{k-1}$. $Q_{k-1}$ can be obtained by numerically integrating the following differential equation from $t_{k-1}$ to $t_k$

$$\frac{dQ}{dt} = \begin{bmatrix} 0_{3\times3} & I_{3\times3} \\ \left(\frac{\partial f}{\partial r}\right)_{3\times3} & 0_{3\times3} \end{bmatrix} Q + Q \begin{bmatrix} 0_{3\times3} & I_{3\times3} \\ \left(\frac{\partial f}{\partial r}\right)_{3\times3} & 0_{3\times3} \end{bmatrix} + \begin{bmatrix} 0_{3\times3} & 0_{3\times3} \\ 0_{3\times3} & \sigma_{w_t}\sigma_{w_t}^T \end{bmatrix} \quad (8)$$

with zero initial values. $\sigma_{w_t}$ is the standard deviation of $w_t$. In the orbit determination process, the matrix differential equations, i.e., Eq. (6) and Eq. (8), are solved in parallel with the integration of the orbit trajectory.

*B. Measurement Equation*

A full-tensor gravity gradiometer is considered in this study to provide gravity gradients for orbit determination. The GRF frame is assumed to be always aligned with the satellite RSW (radial, transverse, and normal) reference frame. The *X*-axis is in the flight direction and is perpendicular to the radius vector (along-track), the *Y*-axis is normal to the orbit plane (cross-track), and the *Z*-axis is radially downwards to the Earth (radial). An onboard star tracker is assumed to simultaneously provide high-precision inertial attitudes, which are used to compute the transformation matrix from ECI to GRF. In addition, the IERS Conventions [19] defines precise models for coordinate transformation from ECI to ECEF. Thus the transformation matrix from ECEF to GRF can be determined directly without a prior knowledge of position and velocity

$$\boldsymbol{C}_e^g = \boldsymbol{C}_i^g \boldsymbol{C}_e^i \tag{9}$$

where the symbols *i*, *e* and *g* denote the frames ECI, ECEF, and GRF, respectively. Referring to Eq. (2), the GGT in GRF can be obtained by

$$\begin{aligned}\boldsymbol{V}_g &= \boldsymbol{C}_e^g \boldsymbol{V}_e \left(\boldsymbol{C}_g^e\right)^T \\ &= \boldsymbol{C}_i^g \boldsymbol{C}_e^i \boldsymbol{V}_e \boldsymbol{C}_i^e \boldsymbol{C}_g^i\end{aligned} \tag{10}$$

where $\boldsymbol{V}_e$ is the GGT in ECEF and can be calculated using the formulas in Appendix A.

The gravity gradients measured by the gradiometer are output as a column vector

$$\vec{V}_g = \begin{bmatrix} V_{g,xx} \\ V_{g,yy} \\ V_{g,zz} \\ \tfrac{1}{2}(V_{g,xy}+V_{g,yx}) \\ \tfrac{1}{2}(V_{g,xz}+V_{g,zx}) \\ \tfrac{1}{2}(V_{g,yz}+V_{g,zy}) \end{bmatrix} \tag{11}$$

Thus the measurement equation is expressed as

$$\boldsymbol{z}_k = \left(\vec{V}_g\right)_k + \boldsymbol{\xi}_k = \boldsymbol{h}(\boldsymbol{x}_k, t_k) + \boldsymbol{\xi}_k \tag{12}$$

where $\boldsymbol{z}_k$ is the 6-dimensional measurement vector at time $t_k$, $\boldsymbol{h}$ is a 6-dimensional vector function, and $\boldsymbol{\xi}_k$ is the observation error. The covariance matrix of $\boldsymbol{\xi}_k$ is called the measurement noise matrix, denoted as $\boldsymbol{R}_k$. The measurement equation considering biases is further given in Eq. (27).

The first-order partial derivative of $\left(\vec{V}_g\right)_k$ with respect to $\boldsymbol{x}_k$ is called the measurement partial matrix and has the following structure

$$(\boldsymbol{H}_k)_{6\times 6} = \frac{\partial \boldsymbol{z}_k}{\partial \boldsymbol{x}_k} = \left[ (\boldsymbol{H}_{r,k})_{6\times 3} \quad (\boldsymbol{H}_{v,k})_{6\times 3} \right] \quad (13)$$

Since the expression of gravity gradients does not explicitly contain velocity, the partials of $(\vec{V}_g)_k$ with respect to the velocity are always set to be zero ($\boldsymbol{H}_{v,k} = \boldsymbol{0}$). Let $\vec{V}_e$ denote the vector form of $V_e$. Let $\boldsymbol{T}_e$ denote the partial derivatives of $\vec{V}_e$ with respect to the position in ECEF. The partials of $(\vec{V}_g)_k$ with respect to the position $r_k$ are

$$(\boldsymbol{H}_{r,k})_{6\times 3} = (\boldsymbol{\Pi}_e^g)_{6\times 6} (\boldsymbol{T}_e)_{6\times 3} (\boldsymbol{C}_i^e)_{3\times 3} \quad (14)$$

with

$$\boldsymbol{\Pi}_e^g = \begin{bmatrix} c_{11}^2 & c_{12}^2 & c_{13}^2 & 2c_{11}c_{12} & 2c_{11}c_{13} & 2c_{12}c_{13} \\ c_{21}^2 & c_{22}^2 & c_{23}^2 & 2c_{21}c_{22} & 2c_{21}c_{23} & 2c_{22}c_{23} \\ c_{31}^2 & c_{32}^2 & c_{33}^2 & 2c_{31}c_{32} & 2c_{31}c_{33} & 2c_{32}c_{33} \\ c_{11}c_{21} & c_{12}c_{22} & c_{13}c_{23} & c_{12}c_{21}+c_{11}c_{22} & c_{13}c_{21}+c_{11}c_{23} & c_{13}c_{22}+c_{12}c_{23} \\ c_{11}c_{31} & c_{12}c_{32} & c_{13}c_{33} & c_{12}c_{31}+c_{11}c_{32} & c_{13}c_{31}+c_{11}c_{33} & c_{13}c_{32}+c_{12}c_{33} \\ c_{21}c_{31} & c_{22}c_{32} & c_{23}c_{33} & c_{22}c_{31}+c_{21}c_{32} & c_{23}c_{31}+c_{21}c_{33} & c_{23}c_{32}+c_{22}c_{33} \end{bmatrix}$$

$$\boldsymbol{T}_e = \begin{bmatrix} \frac{\partial V_{e,xx}}{\partial x} & \frac{\partial V_{e,xx}}{\partial y} & \frac{\partial V_{e,xx}}{\partial z} \\ \frac{\partial V_{e,yy}}{\partial x} & \frac{\partial V_{e,yy}}{\partial y} & \frac{\partial V_{e,yy}}{\partial z} \\ \frac{\partial V_{e,zz}}{\partial x} & \frac{\partial V_{e,zz}}{\partial y} & \frac{\partial V_{e,zz}}{\partial z} \\ \frac{\partial V_{e,xy}}{\partial x} & \frac{\partial V_{e,xy}}{\partial y} & \frac{\partial V_{e,xy}}{\partial z} \\ \frac{\partial V_{e,xz}}{\partial x} & \frac{\partial V_{e,xz}}{\partial y} & \frac{\partial V_{e,xz}}{\partial z} \\ \frac{\partial V_{e,yz}}{\partial x} & \frac{\partial V_{e,yz}}{\partial y} & \frac{\partial V_{e,yz}}{\partial z} \end{bmatrix}$$

where $\boldsymbol{\Pi}_e^g$ is the transformation matrix from $\vec{V}_e$ to $\vec{V}_g$, $\{c_{ij}, i,j=1,2,3\}$ are the elements of $\boldsymbol{C}_e^g$, and $\boldsymbol{T}_e$ comprises the spatial derivatives of the gravity gradients.

The observation error $\boldsymbol{\xi}_k$ includes the GGT transformation error $\boldsymbol{\xi}_k^t$ and the gradiometer noise $\boldsymbol{\xi}_k^n$. The GGT transformation error depends on the accuracies of the attitude measurements and the IERS models. From recent geodetic reports, the relative accuracy of the IERS Earth orientation model is better than $10^{-8}$ [19]. According to the analysis in Chen [16], the corresponding GGT transformation error is smaller than 0.1 mE (1 mE = $10^{-3}$ E) and can be neglected. By contrast, the effect of attitude errors is much more significant. An attitude error of 1 arcsec can cause a GGT transformation error of about 0.01 E at 300 km height. The vector-valued GGT transformation error can be written as

$$\boldsymbol{\xi}_k^t = \boldsymbol{M}_k \begin{bmatrix} \Delta\theta_1 \\ \Delta\theta_2 \\ \Delta\theta_3 \end{bmatrix} \quad (15)$$

where $\Delta\theta_1$, $\Delta\theta_2$, and $\Delta\theta_3$ are Euler angle errors and $M_k$ is a 6 × 3 mapping matrix. The derivation of Eq. (15) is found in Appendix B. Suppose that the Euler angle errors are white Gaussian noises and the noise standard deviation is denoted as $\sigma_\theta$. The covariance of $\xi_k^t$ can be given by

$$R_{\xi_k^t} = M_k M_k^T \sigma_\theta^2 \tag{16}$$

The gradiometer noise is a function of the accelerometer noise. Take the electrostatic gravity gradiometer for example. Three pairs of accelerometers are mounted at the ends of three orthogonal baselines [20]. Specific forces measured by each pair of accelerometers are differenced to provide gravity gradients along that direction. Suppose that the six accelerometers are identical and the noise standard deviation is denoted as $\sigma_0$. Since the accelerometer noises are mutually independent, the covariance of the gradiometer noise $\xi_k^n$ is a diagonal matrix

$$R_{\xi_k^n} = \begin{bmatrix} 2 & 0 & 0 & 0 & 0 & 0 \\ 0 & 2 & 0 & 0 & 0 & 0 \\ 0 & 0 & 2 & 0 & 0 & 0 \\ 0 & 0 & 0 & 1 & 0 & 0 \\ 0 & 0 & 0 & 0 & 1 & 0 \\ 0 & 0 & 0 & 0 & 0 & 1 \end{bmatrix} \frac{\sigma_0^2}{l^2} \tag{17}$$

where $l$ is the length of the gradiometer baseline. $\sqrt{2}\sigma_0/l$ is viewed as the equivalent gradiometer noise level.

The measurement noise matrix is finally obtained by

$$R_k = R_{\xi_k^t} + R_{\xi_k^n} \tag{18}$$

*C. Extended Kalman Filter*

The extended Kalman filter is a linear minimum mean square error (LMMSE) estimator which is based on linearization of nonlinear systems and consists of a recursive calculation of the approximate conditional mean and covariance of the state. The linearization of the dynamic and measurement equation is implemented by the Taylor series expansion in which the second- and higher-order terms are neglected, yielding a first-order EKF.

The recursive estimation algorithm of the EKF is given as follows

$$\bar{x}_k = \varphi(\hat{x}_{k-1}, t_k, t_{k-1}) \tag{19}$$

$$\bar{P}_k = \Phi(t_k, t_{k-1})\hat{P}_{k-1}\Phi(t_k, t_{k-1})^T + Q_{k-1} \tag{20}$$

$$S_k = H_k \bar{P}_k H_k^T + R_k \tag{21}$$

$$K_k = \bar{P}_k H_k S_k^{-1} \tag{22}$$

$$\hat{x}_k = \bar{x}_k + K_k \left[ z_k - h(\bar{x}_k, t_k) \right] \tag{23}$$

$$\hat{P}_k = (I - K_k H_k) \bar{P}_k (I - K_k H_k)^T + K_k R_k K_k^T \tag{24}$$

where $\hat{x}_k$ and $\hat{P}_k$ are the estimated conditional mean and covariance at $t_k$, $\bar{x}_k$ is the predicted state to $t_k$ from $t_{k-1}$, $\bar{P}_k$ and $S_k$ are the state and measurement prediction covariances, and $K_k$ is the filter gain.

The basic EKF algorithm presented above only deals with the case in which the observation errors are additive white noises. If unknown biases are contained in the observations, an augmented state extended Kalman filter (ASEKF) needs to be developed in order to estimate the biases. The augmented state vector consists of the orbital states as well as the biases

$$x_k^a = \begin{bmatrix} x_k \\ B_k \end{bmatrix} \tag{25}$$

where $x_k^a$ is the augmented state vector, $B_k$ is a 6-dimensional bias vector. The dynamic evolution of the bias is modeled as a random walk

$$B_k = B_{k-1} + w_{B,k-1} \tag{26}$$

where $w_{B,k-1}$ is the bias process noise and its covariance $Q_B$ is determined based on the characteristics of the bias. Even for estimating constants, a small fictitious process noise has to be used to prevent divergence of the filter.

The measurement equation for the ASEKF is rewritten as

$$z_k = (\vec{V}_g)_k + B_k + \xi_k = h(x_k, B_k, t_k) + \xi_k \tag{27}$$

And the augmented measurement partial matrix becomes

$$(H_k^a)_{6\times 12} = \frac{\partial z_k}{\partial x_k^a} = \left[ (H_{r,k})_{6\times 3} \quad 0_{6\times 3} \quad I_{6\times 6} \right] \tag{28}$$

where the identity matrix $I_{6\times 6}$ represents the partials of $z_k$ with respect to $B_k$.

The recursive estimation algorithm for the ASEKF can be obtained by replacing $x_k$ in Eqs. (19)-(24) with the augmented vector $x_k^a$. In addition, to balance the matrix $H_k^a$, the algorithm uses mE as the unit of gravity gradients.

*D. Model Configuration*

The model configuration of the EKF includes the definition of deterministic forces for orbital motion and the choice of gravity models for GGT computation. The deterministic forces determine the accuracy of the orbital dynamic model. Its configuration should take into account the computational efficiency as well as the accuracy of the measurement system. For the basic EKF, only the central force and the $J_2$ perturbation are considered. The analytical expression of $f$ is

$$f(r, v, t) = -\frac{GM}{r^3} r \begin{bmatrix} 1 - \frac{3}{2} J_2 \left(\frac{R_E}{r}\right)^2 \left(5\sin^2\phi - 1\right) \\ 1 - \frac{3}{2} J_2 \left(\frac{R_E}{r}\right)^2 \left(5\sin^2\phi - 1\right) \\ 1 - \frac{3}{2} J_2 \left(\frac{R_E}{r}\right)^2 \left(5\sin^2\phi - 3\right) \end{bmatrix} \quad (29)$$

where $J_2$ is the second zonal harmonic coefficient.

For the ASEKF processing model, the gravitational forces up to degree 20 and order 20 are used. The accelerations due to higher degree potential coefficients, lunar and solar gravitational attractions, and the non-gravitational forces (atmospheric drag and solar radiation pressure) are included into the process noise $w_t$. The standard deviation of $w_t$ is required to appropriately reflect the accuracy of the dynamic model. Numerical simulations have been conducted to evaluate the model accuracy. It is found that standard deviations of 0.01 m/s² and 5 × 10⁻⁴ m/s² can be adopted in the basic EKF and the ASEKF, respectively. The accuracy requirement for the state transition matrix evaluation is not stringent. Only the central and $J_2$ terms are considered for both the two filters.

The choice of gravity models for GGT computation depends on the noise level of gradiometers. A 120th degree and order EGM2008 model is used in this study. As shown in Fig. 1, the mean GGT error of this model is on the order of 0.001 E, which is smaller than the noises of most current gradiometers (ranging from 0.01 E to 10 E). Thus, the GGT model errors can be neglected in the filter design. The accuracy requirement for the measurement partial matrix evaluation is also not stringent and only the central and $J_2$ terms are considered.

## IV. Filter Performance Evaluation Metrics

Several criteria are used to evaluate the performance of the orbit determination filter. The first consideration is the filter accuracy. The estimated states can be compared with true values or more accurate values to obtain time-domain error curves, which imply the actual accuracy of the filter. The conditional covariance matrix measures the goodness of the estimate in a probabilistic sense and represents the predicted accuracy of the filter. In this study, the orbital states are estimated in the ECI frame, whereas the position and velocity errors and their predicted variances are expressed in the RSW frame. The total 3-dimensional (3D) position and velocities errors are also used to compare the orbit determination accuracy between different simulation cases.

The filter consistency is another important metric for performance [21]. The consistency consists of two conditions

$$E(x_k - \hat{x}_k) = 0 \quad (30)$$

$$E\left[(x_k - \hat{x}_k)(x_k - \hat{x}_k)\right] = \hat{P}_k \quad (31)$$

Eq. (30) is the unbiasedness requirement. Eq. (31) is the covariance matching requirement, which means that the actual errors should match the filter predicted variances. The consistency is evaluated by the chi-square test. Define the normalized estimation error squared (NEES) as

$$\eta(k) = (x_k - \hat{x}_k)^T \hat{P}_k^{-1} (x_k - \hat{x}_k) \tag{32}$$

Under hypothesis $H_0$ that the EKF is consistent, $\eta(k)$ is chi-square distributed with $n_x$ degrees of freedom

$$\eta(k) \sim \chi^2_{n_x} \tag{33}$$

where $n_x$ is the dimension of the state vector $x$. $H_0$ is accepted if $\eta(k)$ does not exceed the upper limit of a probability region

$$\eta(k) < \chi^2_{n_x}(1-\alpha) \tag{34}$$

where $\alpha$ is the significance level and is set to 5% in this study.

Observability can be inferred from the predicted variances of filter. The observability condition guarantees a "steady flow" of information about the state components and leads to steady-state errors. The steady-state variances are not unique and are affected by many factors, such as data sampling rates and measurement noise levels. The goodness of filter observability can also be inferred from the convergence speed of the filter, i.e., the time of convergence to the steady-state errors.

## V. SIMULATION RESULTS AND DISCUSSIONS

Numerical simulations have been conducted to test the performance of the EKF for GGT-based autonomous orbit determination. The basic EKF cases cover a 6-hour data arc starting from October 1, 2014, 12:00:00.0 (UTC), whereas the ASEKF case covers a 40-hour data arc in order to observe the convergence of the biases. The true orbit ephemerides are generated using a high-precision numerical orbit simulator. The Adams-Bashforth-Moulton method for numerical integration, the EGM2008 model truncated at the 120th degree and order for non-spherical gravitational perturbation, the NRLMSISE-00 model for the atmospheric density, and the analytical formulas for the lunar and solar ephemerides are used in the simulator. The true gravity gradients are generated using a 300th degree and order EGM2008 model. Attitude errors and gradiometer noises at different levels are added to simulate the noisy measurements. In addition, constant biases are also simulated for the ASEKF test.

In the baseline case a nearly circular orbit having a height of 300 km is assumed. The initial osculating orbital elements are listed in Table I. The GGT measurements are simulated using a data-sampling period of 30 s. The standard deviations of the Euler angle error and the gradiometer noise are 10 arcsec and 0.1 E, respectively. Initial errors of 10 km and 10 m/s are added to each component of the ECI position and velocity vectors. The basic EKF algorithm has been implemented for orbit determination. The position and velocity errors in the RSW coordinates are plotted in Fig. 2 and Fig. 3, respectively. The solid lines are the actual errors and the dotted lines are the corresponding plus and minus $3\sigma$ boundaries. As seen from the plots, the filter is consistent: the estimation error has a zero mean and the predicted variances match the actual errors. The filter consistency can also be indicated from the chi-square test results. The NEES values at each epoch are shown in Fig. 4. The upper limit of the probability region is 12.6 for the basic EKF since $n_x = 6$. It is seen that all of the points are inside the region.

Orbit observability can be seen from the predicted variances of the estimation errors. The position errors decrease from 10 km to hundreds of meters level at the first epoch and rapidly ($t < 30$ min) converge to steady states ($1\sigma$ variance) of 31.5-32.5 m (radial), 72.0-95.5 m (along-track), 104-119 m (cross-track), and 142-148 m (3D). The Root Mean Square (RMS) values of the steady-state actual position errors are 29.3 m (radial), 74.8 m (along-track), 89.2 m (cross-track), and 120 m (3D). The position error reduction at the first epoch is due to the position fixing ability of full-tensor GGTs. The oscillation of the steady-state variances is resulted from the combined effects of the satellite rotation motion and the attitude errors. The oscillation frequency is twice that of the rotation motion. In addition, the steady-state position error along the radial direction is much smaller than those along the other two directions. The reason is that the sensitivity factor for the vertical position is much smaller [16]. The velocity errors converge more rapidly ($t < 20$ min) to steady states ($1\sigma$ variance) of 0.172-0.178 m/s (radial), 0.206-0.220 m/s (along-track), 0.241-0.246 m/s (cross-track), and 0.364-0.373 m/s (3D). The RMS values of the steady-state actual velocity errors are 0.111 m/s (radial), 0.099 m/s (along-track), 0.120 m/s (cross-track), and 0.192 m/s (3D). The baseline case demonstrates the feasibility of the EKF for autonomous orbit determination from unbiased gravity gradient measurements. The position accuracy is competitive with that of traditional ground-based station tracking considering the fast convergence of the filter.

The position error of the EKF baseline case can also be compared with the results obtained from the eigendecomposition method given in Chen [16]. Under the similar simulation conditions, the mean position error of the eigendecomposition method is 421 m, which is on the same order of magnitude as the steady-state variance of the 3D position error. However, the EKF performs better than the GGT inversion method which used the $J_2$ gravity model to extract position. The improvement of the position accuracy is attributed to the incorporation of orbital motion and the use of a higher degree reference gravity model.

TABLE I
INITIAL ORBITAL ELEMENTS OF THE REFERENCE ORBIT IN THE BASELINE CASE

| Orbit Element | Initial Conditions |
|---|---|
| height | 300 km |
| eccentricity | 0 |
| inclination | 60 deg |
| right ascension of the ascending node | 120 deg |
| argument of perigee | 0 deg |
| true anomaly | 80 deg |

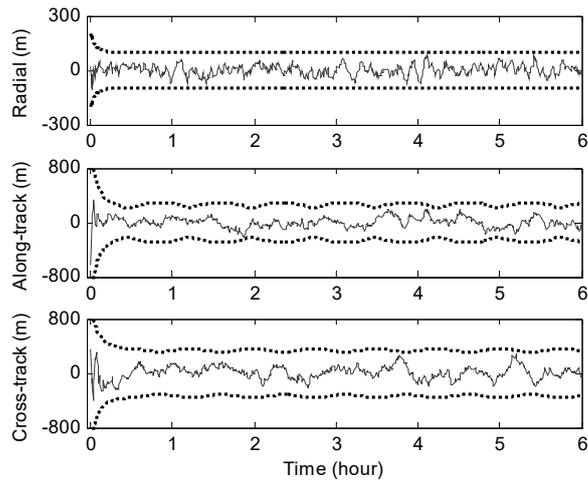

Fig. 2. Radial, along-track, and cross-track position errors and the corresponding 3$\sigma$ boundaries for the baseline case.

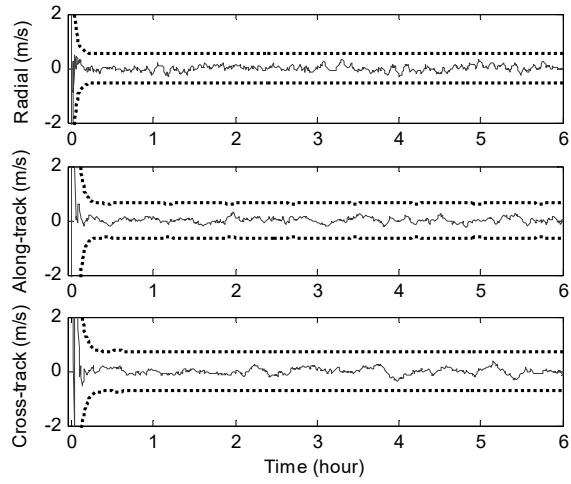

Fig. 3. Radial, along-track, and cross-track velocity errors and the corresponding 3$\sigma$ boundaries for the baseline case.

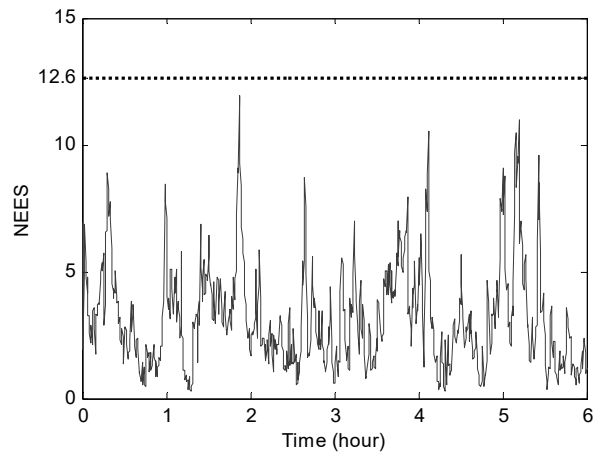

Fig. 4. Normalized state estimation error squared (NEES) and the 95% probability region for the baseline case.

*A. Effects of Initial Errors and Data-Sampling Periods*

Effects of the initial position and velocity errors are tested by comparing the baseline case with another two cases in which larger initial errors of [100 km, 100 km, 100 km, 100 m/s, 100 m/s, 100 m/s] and [1000 km, 1000 km, 1000 km, 1 km/s, 1 km/s, 1 km/s] are assumed, respectively. The other simulation conditions are set to be the same with the baseline case. The evolution of the total 3D position errors is depicted in Fig. 5. Increasing the initial errors slows down the convergence speed of the filter. From Fig.5, the convergence time is 40 min for the 100 km and 100 m/s case, whereas for the 1000 km and 1 km/s case the convergence time is 80 min. In these two cases the filter consistency at the transient stage is not good. That is because the large initial errors increase the nonlinearity and the corrections at the first few epochs are too small. Nevertheless, the steady-state estimation errors are not affected by the initial errors, revealing that the EKF is asymptotically consistent.

Cases having different data-sampling periods are also simulated. The comparison results are summarized in Table II. It is found that increasing the sampling periods degrades the orbit determination accuracy. From Table II, the total position error of the 300-s sampling interval case is almost three times that of the 10-s interval case. The reason is that the observation information collected over the same time span gets less when using a larger sampling period. In addition, the weight of the orbital motion information in the EKF is also reduced, leading to white noise-like estimation errors. An even larger sampling interval of 1000 s has also been tested and the EKF filter still converges. It is shown that the navigation ability under the condition of sparse gravity gradient measurements can be guaranteed. However, an extremely large sampling period will cause divergence of the filter, since the state prediction error is positively correlated with the sampling period and might exceeds the linear correction range of the EKF.

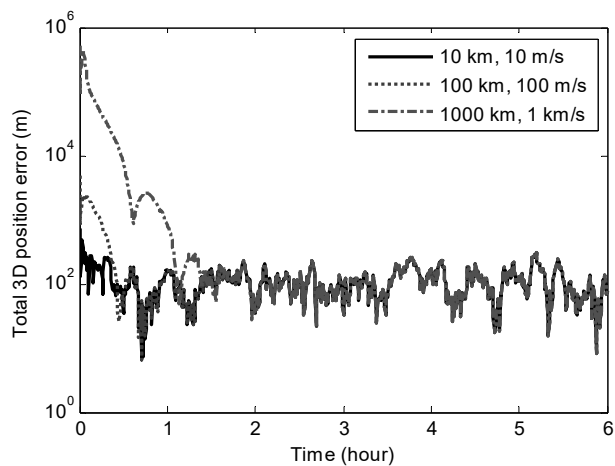

Fig. 5. Evolution of the total 3D position errors for the EKF with different initial estimation errors.

TABLE II
STEADY-STATE VARIANCES FOR THE EKF WITH DIFFERENT DATA-SAMPLING PERIODS

| Sampling Period (s) | Position Variances (1$\sigma$, m) | | | | Velocity Variances (1$\sigma$, m/s) | | | |
|---|---|---|---|---|---|---|---|---|
| | Radial | Along-track | Cross-track | 3D | Radial | Along-track | Cross-track | 3D |
| 10 | 21.4-21.8 | 50.0-67.5 | 71.5-82.5 | 98.5-103 | 0.148-0.152 | 0.185-0.202 | 0.213-0.220 | 0.324-0.332 |
| 60 | 40.0-41.0 | 90.0-117 | 130-147 | 177-184 | 0.190-0.197 | 0.218-0.230 | 0.260-0.265 | 0.392-0.400 |
| 100 | 46.8-47.8 | 106-135 | 152-170 | 206-213 | 0.203-0.210 | 0.226-0.238 | 0.275-0.281 | 0.412-0.422 |
| 300 | 60.4-61.5 | 146-179 | 205-227 | 276-285 | 0.233-0.243 | 0.240-0.251 | 0.316-0.327 | 0.464-0.476 |
| 1000 | 67.2-68.5 | 188-251 | 265-295 | 360-380 | 0.272-0.295 | 0.263-0.275 | 0.390-0.410 | 0.550-0.572 |

*B. Effects of Orbital Heights*

The spherical shape of the Earth's gravitational field indicates that the accuracy of the GGT-based orbit determination is highly dependent of the height and the eccentricity of the orbit but has little relationship with the other orbital elements, such as inclination, true anomaly, etc. Besides the 300 km height case, simulations of circular orbits at heights of 600 km, 1000 km, 2000 km, and 5000 km have also been conducted. The steady-state variances of estimation errors are summarized in Table III. It is seen that the orbit determination accuracy decreases with increasing the orbital height. For the 5000 km case, the steady-state 1$\sigma$ variance of the total 3D position error is 458 m, which is about 3 times that of the 300 km case (148 m). This phenomenon is due to the fact that the GGT signals and their sensitivity factors with respect to position variation are inversely related to the altitude. Thus the GGT-based orbit determination is more suitable for low-Earth orbiting (LEO) satellites. Another phenomenon is that the oscillation of steady-state variances becomes weaker at higher altitudes. The reason is that the effects of attitude errors become weaker at higher altitudes. As seen from Eq. (55), the GGT transformation errors are positively related to gravity gradients, which decreases with height. The filter consistency is not affected by the orbital height. All the NEES values of the four cases in Table III are below the upper limit of 12.6.

TABLE III
STEADY-STATE VARIANCES FOR THE EKF WITH DIFFERENT ORBITAL HEIGHTS

| Orbital Height (km) | Position Variances (1$\sigma$, m) | | | | Velocity Variances (1$\sigma$, m/s) | | | |
|---|---|---|---|---|---|---|---|---|
| | Radial | Along-track | Cross-track | 3D | Radial | Along-track | Cross-track | 3D |
| 600 | 36.4-37.2 | 79.0-102 | 110-125 | 152-158 | 0.180-0.185 | 0.214-0.225 | 0.246-0.250 | 0.375-0.383 |
| 1000 | 43.1-43.8 | 90.0-111 | 120-135 | 167-172 | 0.190-0.195 | 0.222-0.235 | 0.252-0.258 | 0.390-0.396 |
| 2000 | 64.0-64.5 | 122-141 | 149-164 | 214-219 | 0.218-0.221 | 0.245-0.256 | 0.271-0.276 | 0.430-0.434 |
| 5000 | 164-165.8 | 284-295 | 306-315 | 455-458 | 0.303-0.305 | 0.324-0.327 | 0.342-0.345 | 0.562-0.564 |

The effects of dynamic variation of orbital heights are also investigated by assuming an elliptical orbit (eccentricity = 0.26) having a perigee height of 300 km and an apogee height of 5000 km. The evolution of the total 3D position errors and the corresponding 3$\sigma$ boundaries are plotted in Fig. 6. The 6-hour simulation covers 2.5 orbital periods. The predicted variances are consistent with the actual errors. The varying position errors are mainly due to the varying height of the elliptical orbit. The maximum position variance (1$\sigma$) at the apogee height and the minimum position variance (1$\sigma$) at the perigee height are 448 m and 145 m, respectively. The values are consistent with those of the circular orbits at the same heights listed in Table III.

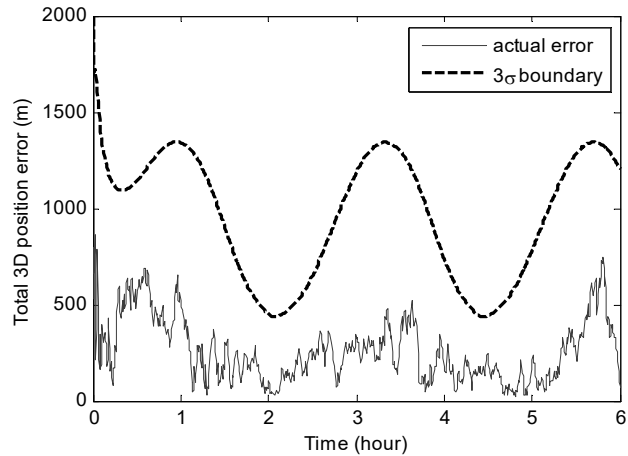

Fig. 6. Evolution of the total 3D position errors and the 3σ boundaries for the elliptical orbit.

*C. Effects of Noise Levels*

The effects of attitude errors are first investigated. A small gradiometer noise of 1 mE is assumed and four different levels of Euler angle error ($\sigma_\theta$), 0.1 arcsec, 1 arcsec, 10 arcsec, 30 arcsec, are tested. The other simulation conditions stay the same with the baseline case. The steady-state variances of estimation errors are summarized in Table IV. It is shown that increasing attitude errors increases not only the total estimation errors but also the oscillation of the steady-state variances. The RMS values of the actual 3D position errors for these four cases are 4.82 m, 20.8 m, 101 m, and 211 m, respectively, and the RMS values of the actual 3D velocity errors are 0.107 m/s, 0.127 m/s, 0.175 m/s, and 0.265 m/s. It is noticed that the radial position accuracy is barely affected by the attitude errors. The reason is that the radial position is mainly related to GGT eigenvalues has little to do with the transformation error [16]. It is noticed that the position accuracy is comparable to the GPS pseudorange performance when the attitude error is reduced to 0.1 arcsec.

To investigate the effects of gradiometer noise on orbit determination accuracy, different levels of gradiometer noise ($\sqrt{2}\sigma_0/l$), 1 mE, 0.01 E, 0.1 E, and 1 E, have been simulated. The 1 E and 0.1 E noise levels represent the precision of most current generation gradiometers, such as the ARKeX's Exploration Gravity Gradiometer and the Gedex's High-Definition Airborne Gravity Gradiometer [14]. The 0.01 E noise level represents the precision of latest gradiometers, such as the GOCE's EGG. The 0.001 E noise level represents the precision of future grade gradiometers [15]. The attitude error is assumed to be 1 arcsec and the other simulation conditions stay the same with the baseline case. The steady-state variances of estimation errors are presented in Table V. The accuracies of position and velocity decrease with increase of gradiometer noise. The RMS values of the actual 3D position errors for these four cases are 20.8 m, 22.1 m, 76.5 m, and 475 m, respectively, and the RMS values of the actual 3D velocity errors are 0.127 m/s, 0.131 m/s, 0.179 m/s, and 0.614 m/s.

TABLE IV
STEADY-STATE VARIANCES FOR THE EKF WITH DIFFERENT LEVELS OF ATTITUDE NOISE

| Attitude Noise (arcsec) | Position Variances ($1\sigma$, m) | | | | Velocity Variances ($1\sigma$, m/s) | | | |
|---|---|---|---|---|---|---|---|---|
| | Radial | Along-track | Cross-track | 3D | Radial | Along-track | Cross-track | 3D |
| 0.1 | 0.640-0.649 | 1.91-2.86 | 2.86-3.46 | 4.00-4.24 | 0.042-0.043 | 0.061-0.071 | 0.071-0.076 | 0.106-0.110 |
| 1 | 0.640-0.650 | 9.25-16.5 | 17.0-20.8 | 22.8-24.5 | 0.043-0.046 | 0.105-0.130 | 0.132-0.140 | 0.180-0.191 |
| 10 | 0.650-0.700 | 51.5-80.0 | 96.0-112 | 123-129 | 0.070-0.097 | 0.178-0.195 | 0.235-0.240 | 0.305-0.321 |
| 30 | 0.650-0.850 | 105-141 | 195-220 | 238-251 | 0.125-0.163 | 0.200-0.208 | 0.301-0.315 | 0.385-0.410 |

TABLE V
STEADY-STATE VARIANCES FOR THE EKF WITH DIFFERENT LEVELS OF GRADIOMETER NOISE

| Gradiometer Noise (E) | Position Variances ($1\sigma$, m) | | | | Velocity Variances ($1\sigma$, m/s) | | | |
|---|---|---|---|---|---|---|---|---|
| | Radial | Along-track | Cross-track | 3D | Radial | Along-track | Cross-track | 3D |
| 0.001 | 0.640-0.650 | 9.25-16.5 | 17.0-20.8 | 22.8-24.5 | 0.043-0.046 | 0.105-0.130 | 0.132-0.140 | 0.180-0.191 |
| 0.01 | 4.83-4.90 | 12.8-18.4 | 18.5-22.0 | 25.9-27.3 | 0.086-0.087 | 0.120-0.135 | 0.136-0.143 | 0.205-0.211 |
| 0.1 | 31.3-31.7 | 57.4-58.5 | 59.6-60.1 | 88.7-89.6 | 0.168-0.169 | 0.194-0.195 | 0.199-0.200 | 0.325-0.326 |
| 1 | 206-209 | 292-296 | 270-272 | 448-452 | 0.451-0.453 | 0.294-0.295 | 0.357-0.358 | 0.646-0.648 |

*D. Effects of Measurement Biases*

Simulated constant biases have been added to true GGT signals to examine the performance of the ASEKF. The biases assumed on the six gravity gradient components are 300 E ($b_{xx}$), -2500 E ($b_{yy}$), 1500 E ($b_{zz}$), 420 E ($b_{xy}$), 900 E ($b_{xz}$), and -120 E ($b_{yz}$), respectively. These significant biases are on the same order of magnitude as the true GGT signals. The initial errors of the bias estimation are set to 10 E. The standard deviation of the fictitious process noise is set to 1 mE. As mentioned earlier, the total simulation time for the ASEKF has been increased to 40 hours. The other simulation conditions are set to be the same with the baseline case.

The position and velocity estimation errors are depicted in Fig. 7 and Fig. 8, respectively. The solid lines are the actual errors and the dotted lines are the corresponding plus and minus $3\sigma$ boundaries. Chi-square test has been conducted to evaluate the filter consistency. The NEES values at each epoch are shown in Fig. 9. The upper limit of the probability region is 21. All the points are inside the region. Out of the 4800 points, only 15 are found outside the region. The radial and cross-track position errors rapidly ($t$ < 3 hours) converge to steady states ($1\sigma$ variance) of 20.0-20.5 m and 30.0-32.0 m, respectively. The along-track position error converges quite slowly and the $1\sigma$ variance at the last epoch is about 400 m. The RMS values of the steady-state actual position errors are 21.5 m (radial), 271 m (along-track), 23.0 m (cross-track), and 273 m (3D). The along-track and cross-track velocity errors rapidly ($t$ < 2 hours) converge to steady states ($1\sigma$ variance) of 0.0198-0.0202 m/s and 0.0355-0.0375 m/s, respectively. The radial velocity error converges quite slowly and $1\sigma$ variance at the last epoch is about 0.47 m/s. The RMS values of the steady-state actual velocity errors are 0.291 m/s (radial), 0.0159 m/s (along-track), 0.0257 m/s (cross-track), and 0.293 m/s (3D). The steady-state variances of the radial and cross-track position components are smaller than those in the baseline case due to the more accurate orbital dynamic model. In addition, the error curves seem much smoother since the orbit prediction constrains the jumps in state estimates.

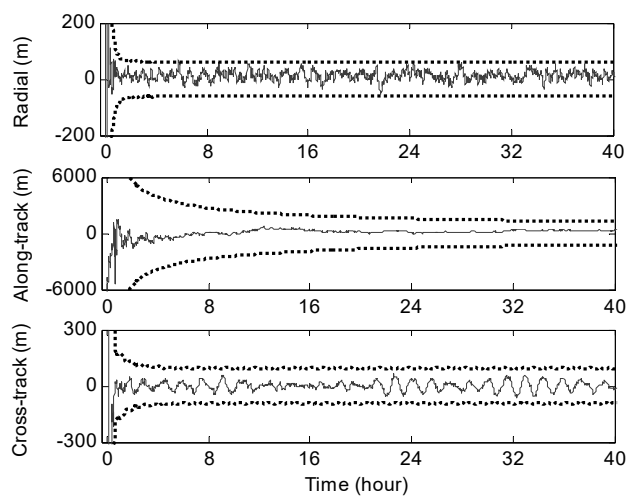

Fig. 7. Radial, along-track, and cross-track position errors and the corresponding 3σ boundaries for the ASEKF case.

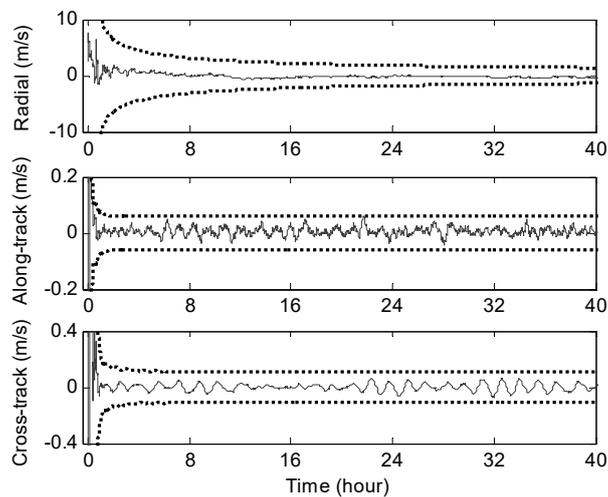

Fig. 8. Radial, along-track, and cross-track velocity errors and the corresponding 3σ boundaries for the ASEKF case.

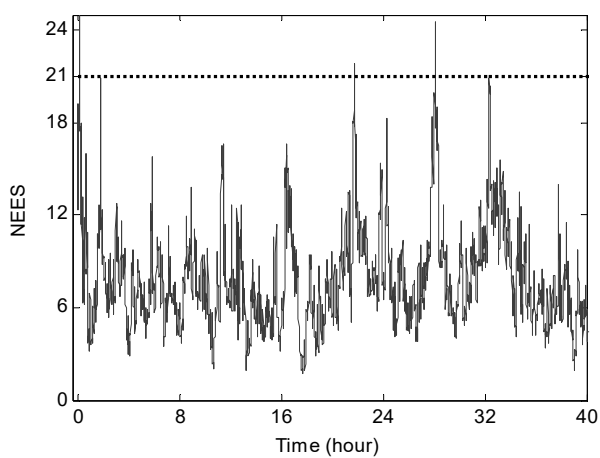

Fig. 9. Normalized state estimation error squared (NEES) as well as the 95% probability region for the ASEKF case.

The evolution of the bias estimation errors and the corresponding $3\sigma$ boundaries is plotted in Fig. 10. Among the six biases, the $b_{xx}$, $b_{yy}$, $b_{zz}$, $b_{xy}$, and $b_{yz}$ components rapidly ($t < 3$ hours) converge to steady states ($1\sigma$ variance) of less than 11 mE, 11 mE, 13 mE, 10 mE, and 16 mE, respectively. However, the $b_{xz}$ component converges quite slowly. The $1\sigma$ variance at the last epoch is about 240 mE. The RMS values of the steady-state actual bias errors are 7.52 mE ($b_{xx}$), 8.74 mE ($b_{yy}$), 7.26 mE ($b_{zz}$), 6.73 mE ($b_{xy}$), 158 mE ($b_{xz}$), and 11.1 mE ($b_{yz}$). The poor observability of $b_{xz}$ accounts for the large errors in the along-track position estimation. As seen from Fig. 7 and Fig. 9, the along-track orbit position error is negatively related to the $b_{xz}$ error, indicating that the ASEKF cannot easily distinguish the variations of these two states. This is due to the fact that the along-track position error varies slowly during the orbit evolution and its dynamic behavior resembles a bias. To illustrate this, assume a Kepler circular orbital motion with an initial error in the true anomaly. The errors of the along-track position as well as the radial velocity both stay constant in the orbit prediction, leading to nearly constant errors of the gravity gradient component $V_{g,xz}$. In addition, the biasedness exists in the $b_{xz}$ estimation, leading to the biasedness in the estimation of along-track position and radial velocity.

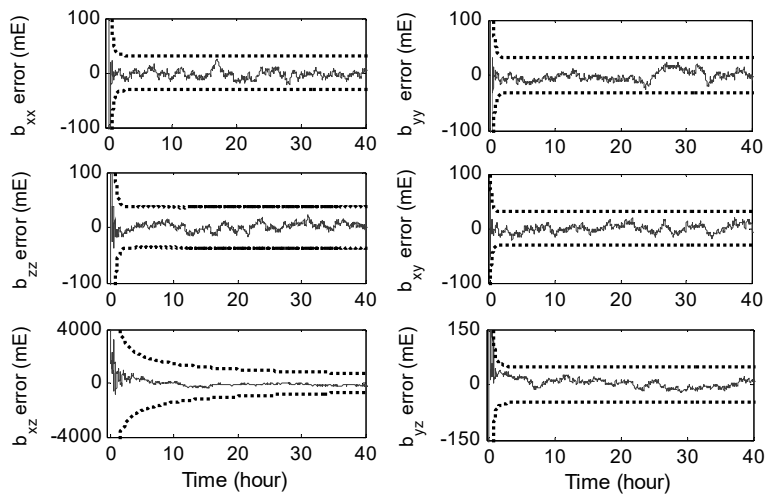

Fig. 10. Evolution of the bias estimation errors and the corresponding $3\sigma$ boundaries for the ASEKF case.

## VI. APPLICATION TO REAL GOCE DATA

Real flight data from the GOCE satellite have been used to test the orbit determination algorithm. The satellite was launched into a sun-synchronous orbit on March 17, 2009, with a mean apogee height of 278.65 km and an inclination of 96.7 deg [22]. The gradiometer onboard the satellite was originally designed to determine the medium- and short-wavelength parts of the gravity field. To ensure a stable and quiet environment, a drag-free control mode was employed. The non-gravitational forces (drag and solar radiation pressure) were continuously compensated by an electric propulsion engine. In addition, three advanced star trackers and two dual-frequency GPS receivers were carried onboard to provide high-precision attitude and orbit information capabilities.

The gravity gradients were sampled at a rate of 1 Hz. The gradiometer achieved the best performance within the MBW from 5

mHz to 0.1 Hz. Data analysis has shown that the noise density level inside the MBW is approximately 20 mE/√Hz for $V_{g,xx}$, $V_{g,yy}$, $V_{g,zz}$ and $V_{g,xz}$, 350 mE/√Hz for $V_{g,xy}$, and 500 mE/√Hz for $V_{g,yz}$. The larger noise levels of $V_{g,xy}$ and $V_{g,yz}$ are due to the less sensitive axes of the accelerometers. Below the MBW the noises increase with $1/f$ and are superimposed by cyclic distortions [23]. Therefore, the biases of GOCE gradiometer are not constant. They are slowly drifting and have small-amplitude oscillations. To cope with the time-varying biases, a large process noise is assumed. The standard deviation is set to 20 mE.

A 40-hour data arc starting from September 8, 2013, 00:00:00.0 (GPS time) are used for the test. The data arc covers about 27 orbital periods and includes 4800 measurement epochs with a data-sampling period of 30 s. The GOCE Level 1b product EGG_NOM_1b which contains raw GGT measurements (EGG_GGT) and gradiometer inertial attitudes (EGG_IAQ) are used as inputs. The outliers in the raw GGTs have been detected and removed via polynomial fitting. The standard deviation of the attitude errors is set to 2 arcsec [24]. To assess the accuracy of the orbit determination results, the GPS-derived high-precision orbits with an accuracy of 2 cm are used as references. The reference values of the GGT biases have also been obtained through comparing the actual measurements with GGTs calculated from the 300th degree and order EGM2008 model.

The ASEKF has been implemented for orbit determination. The initial position and velocity errors are 10 km and 10 m/s. The initial bias errors are 10 E. The position and velocity errors are shown in Fig. 11 and Fig. 12, respectively. The chi-square test has been conducted to evaluate the filter consistency. The NEES values are shown in Fig. 13. Only 22 points are found outside the probability region. Similar to the ASEKF simulation case, the radial and cross-track position errors converge rapidly ($t < 4$ hours) to steady states ($1\sigma$ variance) of 60-66 m and 55-75 m, respectively. The along-track position error converges slowly and $1\sigma$ variance at the last epoch is about 900 m. The RMS values of the steady-state actual position errors are 15.6 m (radial), 559 m (along-track), 52.2 m (cross-track), and 562 m (3D). The along-track and cross-track velocity errors converge rapidly ($t < 5$ hours) to steady states ($1\sigma$ variance) of 0.050-0.052 m/s and 0.065-0.080 m/s, respectively, whereas the radial velocity error converges slowly with a final $1\sigma$ variance value of about 1.05 m/s. The RMS values of the steady-state actual velocity errors are 0.658 m/s (radial), 0.0166 m/s (along-track), 0.0587 m/s (cross-track), and 0.662 m/s (3D). The GGT biases have also been estimated along with the orbits and the errors are depicted in Fig. 14. The $b_{xx}$, $b_{yy}$, $b_{zz}$, $b_{xy}$, and $b_{yz}$ components rapidly ($t < 3$ hours) converge to steady states ($1\sigma$ variance) of less than 43 mE, 42 mE, 80 mE, 83 mE, and 105 mE, respectively. The $b_{xz}$ component converges very slowly with a final $1\sigma$ variance value of about 600 mE. The RMS values of the steady-state actual bias errors are 12.3 mE ($b_{xx}$), 30.7 mE ($b_{yy}$), 21.4 mE ($b_{zz}$), 85.5 mE ($b_{xy}$), 351 mE ($b_{xz}$), and 101 mE ($b_{yz}$). The use of bias process noise allows successful estimation of the varying biases. The reference values as well as the estimated values of $b_{yy}$ are plotted in Fig. 15 as an example. It is seen that the estimates follow closely the references values and compensate the drift as well as the small periodic fluctuations. Although not shown in a figure, the post-fit measurement residuals have been calculated. The statistical standard deviations of the six components are 12 mE, 11 mE, 16 mE, 335 mE, 15 mE, and 494 mE, respectively, consistent with the flat noise levels in the MBW.

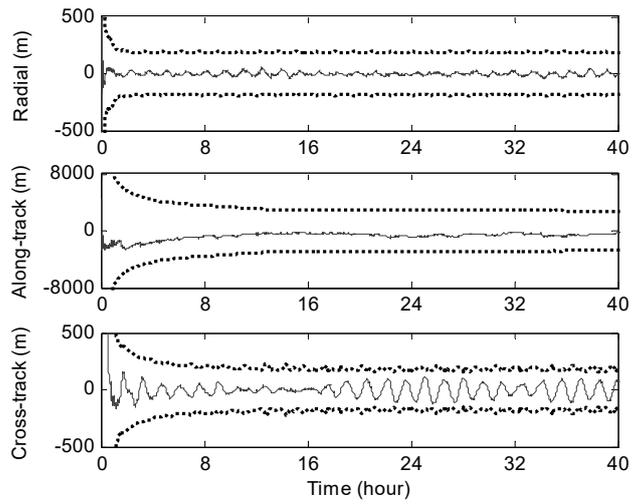

Fig. 11. Radial, along-track, and cross-track position errors and the corresponding $3\sigma$ boundaries for the GOCE case.

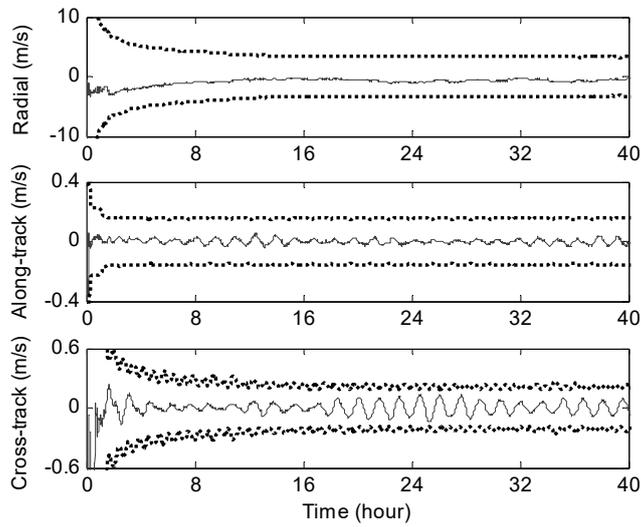

Fig. 12. Radial, along-track, and cross-track velocity errors and the corresponding $3\sigma$ boundaries for the GOCE case.

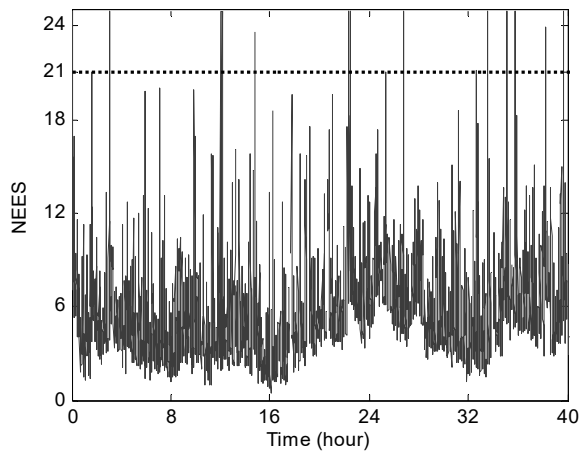

Fig. 13. Normalized state estimation error squared (NEES) as well as the 95% probability region for the GOCE case.

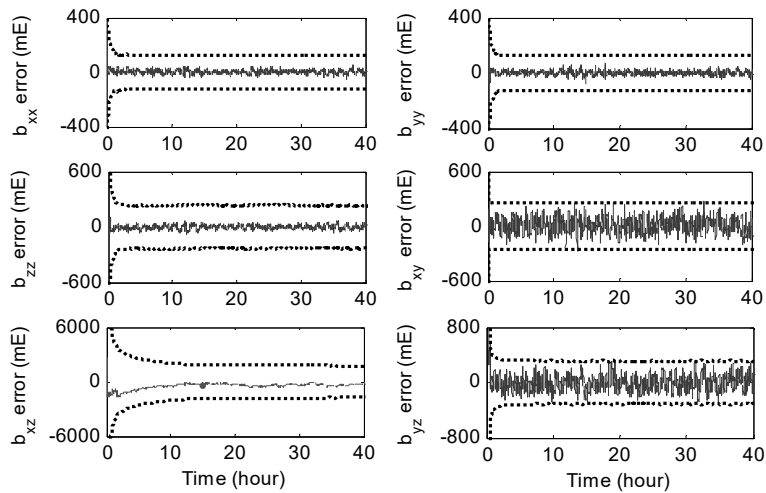

Fig. 14. Evolution of the bias estimation errors and the corresponding 3$\sigma$ boundaries for the GOCE case.

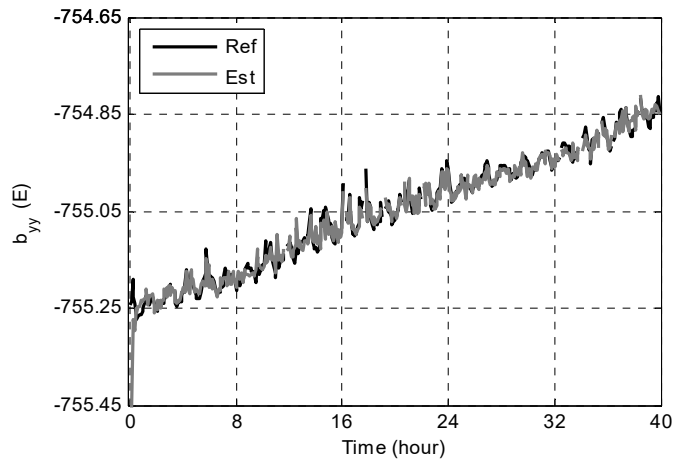

Fig. 15. The estimated and reference values of the $b_{yy}$ bias of GOCE.

## VII. Conclusions

A geophysical navigation scheme using the Earth's gravitational field is proposed in this paper for satellite autonomous orbit determination. The gravity gradiometer which measures full-tensor gravity gradients is surveyed as a primary sensor. High-precision attitude sensors are assumed available to remove the orientation contributions in the observations. The novel method does not rely on ground stations or any other satellites and is suitable for fully autonomous navigation.

The Extended Kalman filter is implemented for orbit determination. Several simulations are conducted to test the algorithm. The initial errors only affect the convergence speed but not the steady-state errors. Increasing the sampling periods degrades the filter accuracy. In addition, the orbit determination accuracy is affected by the orbital heights as well as the attitude errors and gradiometer noises. With a future grade gradiometer, accuracy comparable to the GPS pseudorange will be possible. The bias on the along-track-radial gravity gradient components worsens the system's observability and leads to large estimation errors of the along-track position and the radial velocity. The along-track position error after 40 hours is about 350m for the real GOCE case. Additional calibrations could be helpful to improve the orbit accuracy. Nevertheless, this study demonstrates the feasibility of the EKF algorithm for GGT-based orbit determination, which could be used as a backup for near Earth autonomous spacecraft operation, or as a primary navigation system for future planetary exploration.

APPENDIX A. CALCULATION OF GGT IN ECEF

The computational form of the gravitation potential is given as follows

$$U = \frac{GM}{R_E} \sum_{n=0}^{\infty} \sum_{m=0}^{n} \left( \bar{C}_{nm} \bar{V}_{nm} + \bar{S}_{nm} \bar{W}_{nm} \right) \tag{35}$$

with

$$\bar{V}_{nm} = \left(\frac{R_E}{r}\right)^{n+1} \bar{P}_{nm}(\sin\phi)\cos(m\lambda)$$

$$\bar{W}_{nm} = \left(\frac{R_E}{r}\right)^{n+1} \bar{P}_{nm}(\sin\phi)\sin(m\lambda)$$

The recursive calculation of $\bar{V}_{nm}$ and $\bar{W}_{nm}$ is given by

$$\bar{V}_{m,m} = \sqrt{\frac{(2m+1)}{(2-\delta_{0,m-1})m}} \left( \frac{xR_E}{r^2} \bar{V}_{m-1,m-1} - \frac{yR_E}{r^2} \bar{W}_{m-1,m-1} \right) \tag{36}$$

$$\bar{W}_{m,m} = \sqrt{\frac{(2m+1)}{(2-\delta_{0,m-1})m}} \left( \frac{xR_E}{r^2} \bar{W}_{m-1,m-1} + \frac{yR_E}{r^2} \bar{V}_{m-1,m-1} \right) \tag{37}$$

$$\bar{V}_{n,m} = \sqrt{\frac{(2n+1)(2n-1)}{(n+m)(n-m)}} \frac{zR_E}{r^2} \bar{V}_{n-1,m} - \sqrt{\frac{(n+m-1)(n-m-1)(2n+1)}{(n+m)(n-m)(2n-3)}} \left(\frac{R_E}{r}\right)^2 \bar{V}_{n-2,m} \tag{38}$$

$$\bar{W}_{n,m} = \sqrt{\frac{(2n+1)(2n-1)}{(n+m)(n-m)}} \frac{zR_E}{r^2} \bar{W}_{n-1,m} - \sqrt{\frac{(n+m-1)(n-m-1)(2n+1)}{(n+m)(n-m)(2n-3)}} \left(\frac{R_E}{r}\right)^2 \bar{W}_{n-2,m} \tag{39}$$

where $x$, $y$, and $z$ here refer to the three position components in the ECEF frame. The initial values are

$$\bar{V}(0,0) = \frac{R_E}{r}, \bar{W}(0,0) = 0 \tag{40}$$

The diagonal terms $V_{mm}$ and $W_{mm}$ can be first calculated using Eq. (36) and Eq. (37). By fixing $m$, the terms $V_{nm}$ and $W_{nm}$ ($n > m$) can be subsequently obtained using Eq. (38) and Eq. (39).

The gravitational force is obtained by differentiation of Eq. (35) and is given by

$$\boldsymbol{a}_e = \frac{GM}{R_E} \sum_{n=0}^{\infty} \sum_{m=0}^{n} \left( \bar{C}_{nm} \frac{\partial \bar{V}_{nm}}{\partial \boldsymbol{r}_e} + \bar{S}_{nm} \frac{\partial \bar{W}_{nm}}{\partial \boldsymbol{r}_e} \right) \tag{41}$$

where $\boldsymbol{a}_e$ and $\boldsymbol{r}_e$ refer to the gravitational force vector and position vector in the ECEF coordinate frame. The partial derivatives of $V_{nm}$ and $W_{nm}$ are

$$\frac{\partial \bar{V}_{nm}}{\partial x} = \begin{cases} -\frac{1}{R_E} \sqrt{\frac{(n+2)(n+1)(2n+1)}{2(2n+3)}} \bar{V}_{n+1,1}, & m=0 \\ \frac{1}{2R_E} \left( -\sqrt{\frac{(n+m+2)(n+m+1)(2n+1)}{(2n+3)}} \bar{V}_{n+1,m+1} + \sqrt{\frac{2(n-m+2)(n-m+1)(2n+1)}{(2-\delta_{0,m-1})(2n+3)}} \bar{V}_{n+1,m-1} \right), & m>0 \end{cases} \tag{42}$$

$$\frac{\partial \bar{V}_{nm}}{\partial y} = \begin{cases} -\frac{1}{R_E} \sqrt{\frac{(n+2)(n+1)(2n+1)}{2(2n+3)}} \bar{W}_{n+1,1}, & m=0 \\ \frac{1}{2R_E} \left( -\sqrt{\frac{(n+m+2)(n+m+1)(2n+1)}{(2n+3)}} \bar{W}_{n+1,m+1} - \sqrt{\frac{2(n-m+2)(n-m+1)(2n+1)}{(2-\delta_{0,m-1})(2n+3)}} \bar{W}_{n+1,m-1} \right), & m>0 \end{cases} \tag{43}$$

$$\frac{\partial \bar{V}_{nm}}{\partial z} = -\frac{1}{R_E} \sqrt{\frac{(n-m+1)(n+m+1)(2n+1)}{(2n+3)}} \bar{V}_{n+1,m} \tag{44}$$

$$\frac{\partial \bar{W}_{nm}}{\partial x} = \begin{cases} 0, & m=0 \\ \frac{1}{2R_E} \left( -\sqrt{\frac{(n+m+2)(n+m+1)(2n+1)}{(2n+3)}} \bar{W}_{n+1,m+1} + \sqrt{\frac{2(n-m+2)(n-m+1)(2n+1)}{(2-\delta_{0,m-1})(2n+3)}} \bar{W}_{n+1,m-1} \right), & m>0 \end{cases} \tag{45}$$

$$\frac{\partial \bar{W}_{nm}}{\partial y} = \begin{cases} 0, & m=0 \\ \frac{1}{2R_E} \left( \sqrt{\frac{(n+m+2)(n+m+1)(2n+1)}{(2n+3)}} \bar{V}_{n+1,m+1} + \sqrt{\frac{2(n-m+2)(n-m+1)(2n+1)}{(2-\delta_{0,m-1})(2n+3)}} \bar{V}_{n+1,m-1} \right), & m>0 \end{cases} \tag{46}$$

$$\frac{\partial \bar{W}_{nm}}{\partial z} = -\frac{1}{R_E} \sqrt{\frac{(n-m+1)(n+m+1)(2n+1)}{(2n+3)}} \bar{W}_{n+1,m} \tag{47}$$

The gravity gradient tensor can be further obtained by differentiation of Eq. (41) and is given by

$$\boldsymbol{V} = \frac{GM}{R_E} \sum_{n=0}^{\infty} \sum_{m=0}^{n} \left( \bar{C}_{nm} \frac{\partial^2 \bar{V}_{nm}}{\partial \boldsymbol{r}_e^2} + \bar{S}_{nm} \frac{\partial^2 \bar{W}_{nm}}{\partial \boldsymbol{r}_e^2} \right) \tag{48}$$

where $\boldsymbol{V}$ represents GGT and should be distinguished from $\bar{V}_{nm}$. The spatial derivatives of $\boldsymbol{V}$ are

$$\frac{\partial V_{ij}}{\partial x} = \frac{GM}{R_E} \sum_{n=0}^{\infty} \sum_{m=0}^{n} \left( \bar{C}_{nm} \frac{\partial^3 \bar{V}_{nm}}{\partial i \partial j \partial x} + \bar{S}_{nm} \frac{\partial^3 \bar{W}_{nm}}{\partial i \partial j \partial x} \right) \tag{49}$$

$$\frac{\partial V_{ij}}{\partial y} = \frac{GM}{R_E} \sum_{n=0}^{\infty} \sum_{m=0}^{n} \left( \bar{C}_{nm} \frac{\partial^3 \bar{V}_{nm}}{\partial i \partial j \partial y} + \bar{S}_{nm} \frac{\partial^3 \bar{W}_{nm}}{\partial i \partial j \partial y} \right) \tag{50}$$

$$\frac{\partial V_{ij}}{\partial z} = \frac{GM}{R_E} \sum_{n=0}^{\infty} \sum_{m=0}^{n} \left( \bar{C}_{nm} \frac{\partial^3 \bar{V}_{nm}}{\partial i \partial j \partial z} + \bar{S}_{nm} \frac{\partial^3 \bar{W}_{nm}}{\partial i \partial j \partial z} \right) \quad (51)$$

where $i$ and $j$ refer to the indexes $x$, $y$, and $z$. The second- and third-order partial derivatives of $V_{nm}$ and $W_{nm}$ can be derived from Eqs. (42)-(47) and are not given here.

APPENDIX B. DERIVATION OF (15)

Let $\Delta C_i^g$ denote the rotation uncertainty of $C_i^g$. The resulting GGT transformation error in the matrix form will be

$$\Delta V_g = \Delta C_i^g C_e^i V_e C_i^e C_g^i + C_i^g C_e^i V_e C_i^e \Delta C_g^i \quad (52)$$

The transformation matrix $C_i^g$ can be expressed as a function of three Euler angles

$$C_i^g = \begin{bmatrix} \cos\theta_3 \cos\theta_1 - \sin\theta_2 \sin\theta_3 \sin\theta_1 & \cos\theta_3 \sin\theta_1 + \sin\theta_2 \sin\theta_3 \cos\theta_1 & -\cos\theta_2 \sin\theta_3 \\ -\cos\theta_2 \sin\theta_1 & \cos\theta_2 \cos\theta_1 & \sin\theta_2 \\ \sin\theta_3 \cos\theta_1 + \sin\theta_2 \cos\theta_3 \sin\theta_1 & \sin\theta_3 \sin\theta_1 - \sin\theta_2 \cos\theta_3 \cos\theta_1 & \cos\theta_2 \cos\theta_3 \end{bmatrix} \quad (53)$$

where $\theta_1$, $\theta_2$, and $\theta_3$ are the yaw, pitch, and roll angles, respectively. The relationship between $\Delta C_i^g$ and the Euler angle errors $\Delta\theta_1$, $\Delta\theta_2$, and $\Delta\theta_3$ can be derived by linearization of Eq. (53)

$$\Delta C_i^g = L_1 \Delta\theta_1 + L_2 \Delta\theta_2 + L_3 \Delta\theta_3 \quad (54)$$

with

$$L_1 = \frac{\partial C_i^g}{\partial \theta_1} = \begin{bmatrix} -\cos\theta_3 \sin\theta_1 - \sin\theta_2 \sin\theta_3 \cos\theta_1 & \cos\theta_3 \cos\theta_1 - \sin\theta_2 \sin\theta_3 \sin\theta_1 & 0 \\ -\cos\theta_2 \cos\theta_1 & -\cos\theta_2 \sin\theta_1 & 0 \\ -\sin\theta_3 \sin\theta_1 + \sin\theta_2 \cos\theta_3 \cos\theta_1 & \sin\theta_3 \cos\theta_1 + \sin\theta_2 \cos\theta_3 \sin\theta_1 & 0 \end{bmatrix}$$

$$L_2 = \frac{\partial C_i^g}{\partial \theta_2} = \begin{bmatrix} -\cos\theta_2 \sin\theta_3 \sin\theta_1 & \cos\theta_2 \sin\theta_3 \cos\theta_1 & \sin\theta_2 \sin\theta_3 \\ \sin\theta_2 \sin\theta_1 & -\sin\theta_2 \cos\theta_1 & \cos\theta_2 \\ \cos\theta_2 \cos\theta_3 \sin\theta_1 & -\cos\theta_2 \cos\theta_3 \cos\theta_1 & -\sin\theta_2 \cos\theta_3 \end{bmatrix}$$

$$L_3 = \frac{\partial C_i^g}{\partial \theta_3} = \begin{bmatrix} -\sin\theta_3 \cos\theta_1 - \sin\theta_2 \cos\theta_3 \sin\theta_1 & -\sin\theta_3 \sin\theta_1 + \sin\theta_2 \cos\theta_3 \cos\theta_1 & -\cos\theta_2 \cos\theta_3 \\ 0 & 0 & 0 \\ \cos\theta_3 \cos\theta_1 - \sin\theta_2 \sin\theta_3 \sin\theta_1 & \cos\theta_3 \sin\theta_1 + \sin\theta_2 \sin\theta_3 \cos\theta_1 & -\cos\theta_2 \sin\theta_3 \end{bmatrix}$$

where $L_1$, $L_2$, and $L_3$ are the coefficient matrices. Therefore, the GGT transformation error in the matrix form becomes

$$\Delta V_g = \left( L_1 C_e^i V_e C_i^e C_g^i + C_i^g C_e^i V_e C_i^e L_1^T \right) \Delta\theta_1 \\ + \left( L_2 C_e^i V_e C_i^e C_g^i + C_i^g C_e^i V_e C_i^e L_2^T \right) \Delta\theta_2 \quad (55) \\ + \left( L_3 C_e^i V_e C_i^e C_g^i + C_i^g C_e^i V_e C_i^e L_3^T \right) \Delta\theta_3$$

Thus the vector-valued GGT transformation error can be written as

$$\xi_k^t = M_k \begin{bmatrix} \Delta\theta_1 \\ \Delta\theta_2 \\ \Delta\theta_3 \end{bmatrix} \tag{56}$$

where the 6 × 3 mapping matrix $M_k$ can be directly obtained from Eq. (55).

## REFERENCES

[1] Rice, H., Kelmenson, S., and Mendelsohn, L., "Geophysical navigation technologies and applications," in *IEEE Position Location and Navigation Symposium*, 2004, pp. 618-624.

[2] Shorshi, G. and Bar-Itzhack, I. Y., "Satellite autonomous navigation based on magnetic field measurements," *J. Guid. Control Dynam.*, vol. 18, no. 4, pp. 843-850, Jul.-Aug. 1995.

[3] Roh, K.M., Park, S.Y., and Choi, K.H., "Orbit determination using the geomagnetic field measurement via the unscented Kalman filter," *J. Guid. Control Dynam.*, vol. 44, no. 1, pp. 246-253, Jan.-Feb. 2007.

[4] Abdelrahman, M. and Park, S. Y., "Sigma-point Kalman filtering for spacecraft attitude and rate estimation using magnetometer measurements," *IEEE Trans. Aero. Elec. Sys.*, vol. 47, no. 2, Apr. 2011.

[5] Soken, H. E. and Hajiyev, C., "UKF-based reconfigurable attitude parameters estimation and magnetometer calibration," *IEEE Trans. Aero. Elec. Sys.*, vol. 48, no. 3, Jul. 2012.

[6] Tippenhauer, N. O., Pöpper, C., Rasmussen, K. B., and Čapkun, S., "On the requirements for successful GPS spoofing attacks," in *ACM Conference on Computer and Communications Security*, Chicago, Illinois, Oct. 17-21, 2011.

[7] Britting, K. R., Madden, Jr., S.J., and Hildebrant, R.A., "The impact of gradiometer techniques on the performance of inertial navigation systems," in *AIAA Guidance and Control Conference*, Stanford, California, Aug. 14-16, 1972, AIAA 72-850.

[8] Grubin, C., "Accuracy improvement in a gravity gradiometer-aided cruise inertial navigator subjected to deflections of the vertical," in *AIAA Guidance and Control Conference*, Boston, Massachusetts, Aug. 20-22, 1975, AIAA 75-1090.

[9] Metzger, E.H. and Jircitano, A., "Inertial navigation performance improvement using gravity gradient matching techniques," in *AIAA Guidance and Control Conference*, Boston, Massachusetts, Aug. 20-22, 1975, AIAA 75-1092.

[10] Affleck, C.A. and Jircitano, A., "Passive gravity gradiometer navigation system," In *IEEE Position Location and Navigation Symposium*, 1990, pp. 60-66.

[11] Gleason, D.M., "Passive airborne navigation and terrain avoidance using gravity gradiometry," *J. Guid. Control Dynam.*, vol. 18, no. 6, pp. 1450-1458, Nov.-Dec. 1995.

[12] Richeson, A., "Gravity gradiometer aided inertial navigation within non-GNSS environments," Ph.D. dissertation, Dept. Aero. Eng., Univ. of Maryland, MD, 2008.

[13] Pavlis, K., Holmes, A., Kenyon, C., and Factor, K., "The development and evaluation of the earth gravitational model 2008 (EGM2008)," *J. Geophys. Res.*, vol. 117, B04406, 2008.

[14] DiFrancesco, D., Grierson, A., Kaputa, D., and Meyer, T., "Gravity gradiometer systems-advances and challenges," *Geophys. Prospect.*, vol. 57, no. 4, pp. 615-623, Jul. 2009.

[15] Carraz, O., Siemes, C., Massotti, L., Haagmans, R., and Silvestrin, P., "A spaceborne gravity gradiometer concept based on cold atom interferometers for measuring Earth's gravity field," *Microgravity Sci. Tec.*, vol. 26, no. 3, pp. 139-145, Oct. 2014.

[16] Chen, P., Sun, X., and Han, C., "Gravity gradient tensor eigendecomposition for spacecraft positioning," *J. Guid. Control Dynam.*, vol. 38, no. 11, pp. 2200-2206, Nov. 2015.

[17] Sun, X., Chen, P., Han, Y., Macabiau, C., and Han, C., "Test of GOCE EGG data for spacecraft positioning," In *Proc. 5th International GOCE User Workshop*, ESA SP-728, Paris, France, Nov. 24-28, 2014.

[18] Hofmann-Wellenhof, B. and Moritz, H., *Physical Geodesy*, Springer-Verlag Wien, 2005, pp. 3-41.

[19] Petit, G. and Luzum, B., "IERS Conventions 2010," IERS Technical Note no.36, Bundesamt für Kartographie und Geodäsie, Frankfurt am Main, Germany, 2010.

[20] Cesare, S., "Performance requirements and budgets for the gradiometric mission," fourth ed., Thales Alenia Space, GO-TN-AI-0027, Feb. 22, 2008.

[21] Bar-Shalom, Y., Li, R., and Kirubarajan, T., "Estimation with Applications to Tracking and Navigation: Theory Algorithms and Software," John Wiley & Sons, Inc. 2001, pp. 222-245.

[22] Floberghagen, R., Fehringer, M., Lamarre, D., Muzi, D., Frommknecht, B., Steiger, C., Piñeiro, J., and Da Costa, A., "Mission design, operation and exploitation of the gravity field and steady-state ocean circulation explorer mission," *J. Geodesy*, vol. 85, no. 11, pp. 749-758, Nov. 2011.

[23] Rummel, R., Yi, W., and Stummer, C., "GOCE gravitational gradiometry," *J. Geodesy*, vol. 85, no. 11, pp. 777-790, Nov. 2011.

[24] Stummer, C., Siemes, C., Pail, R., Frommknecht, B., and Floberghagen, R., "Upgrade of the GOCE Level 1b gradiometer processor," *Adv. Space Res.*, vol. 49, no. 4, pp. 739-752, Feb. 2012.

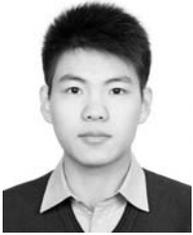

**Xiucong Sun** was born in Laiwu, China, in 1988. He received his B.S. degree in aerospace engineering from Beihang University, Beijing, in 2010.

   He is a Ph.D. student at the school of astronautics, Beihang University. Since 2014, he has been with ENAC (Ecole Nationale de l'Aviation Civile) TELECOM laboratory as a visiting researcher, funded by the China Scholarship Council. The focus of his current research mainly lies in gravity gradient based spacecraft navigation.

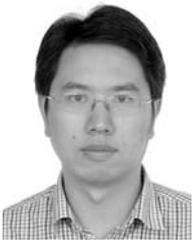

**Pei Chen** was born in Beijing, China, in 1979. He received his Ph.D. degree in aerospace engineering from Beihang University, Beijing, in 2008.

   He is currently an associate professor at the school of astronautics, Beihang University. His current research activities comprise spacecraft navigation, GNSS application, and astrodynamics and simulation.
.

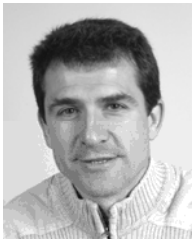

**Christophe Macabiau** graduated as an electronics engineer in 1992 from the ENAC (Ecole Nationale de l'Aviation Civile) in Toulouse, France. He received his Ph.D. in 1997.

   Since 1994 he has been working on the application of satellite navigation techniques to civil aviation and has been in charge of the TELECOM Lab of the ENAC since 2000.

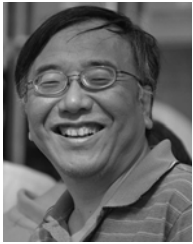

**Chao Han** was born in Beijing, China, in 1960. He received his M.S. and Ph.D. degrees in Applied Mechanics from Beihang University in 1985 and 1989, respectively.

   He is currently a professor at the school of astronautics, Beihang University. He is also an editorial board member of Chinese Journal of Aeronautics. The focus of his research activities lies in the area of spacecraft orbit and attitude dynamics, spacecraft guidance, navigation, and control.

Fig. 1.  The precision of EGM2008 gravity models truncated at different degrees. The benchmark is a 300th degree and order version. The reference height is 300 km.

Fig. 2.  Radial, along-track, and cross-track position errors and the corresponding 3$\sigma$ boundaries for the baseline case.

Fig. 3.  Radial, along-track, and cross-track velocity errors and the corresponding 3$\sigma$ boundaries for the baseline case.

Fig. 4.  Normalized state estimation error squared (NEES) and the 95% probability region for the baseline case.

Fig. 5.  Evolution of the total 3D position errors for the EKF with different initial estimation errors.

Fig. 6.  Evolution of the total 3D position errors and the 3$\sigma$ boundaries for the elliptical orbit.

Fig. 7.  Radial, along-track, and cross-track position errors and the corresponding 3$\sigma$ boundaries for the ASEKF case.

Fig. 8.  Radial, along-track, and cross-track velocity errors and the corresponding 3$\sigma$ boundaries for the ASEKF case.

Fig. 9.  Normalized state estimation error squared (NEES) as well as the 95% probability region for the ASEKF case.

Fig. 10.  Evolution of the bias estimation errors and the corresponding 3$\sigma$ boundaries for the ASEKF case.

Fig. 11.  Radial, along-track, and cross-track position errors and the corresponding 3$\sigma$ boundaries for the GOCE case.

Fig. 12.  Radial, along-track, and cross-track velocity errors and the corresponding 3$\sigma$ boundaries for the GOCE case.

Fig. 13.  Normalized state estimation error squared (NEES) as well as the 95% probability region for the GOCE case.

Fig. 14.  Evolution of the bias estimation errors and the corresponding 3$\sigma$ boundaries for the GOCE case.

Fig. 15.  The estimated and reference values of the $b_{yy}$ bias of GOCE.